Numerical simulation of the nocturnal cooling effect of urban trees considering the leaf area density distribution

Haruki Oshio*, Tomoki Kiyono[1], and Takashi Asawa

Department of Architecture and Building Engineering, School of Environment and Society, Tokyo Institute of Technology, 4259 Nagatsuda-cho, Midori-ku, Yokohama 226-8502, Japan, E-mail address: oshio.h.aa@m.titech.ac.jp (Haruki Oshio); asawa.t.aa@m.titech.ac.jp (Takashi Asawa)

[1] Current affiliation: Earth System Division, National Institute for Environmental Studies, 16-2 Onogawa, Tsukuba-shi, Ibaraki 305-8506, Japan, E-mail address: kiyono.tomoki@nies.go.jp

* Corresponding author. Tel.: +81 459245510; E-mail address: oshio.h.aa@m.titech.ac.jp (Haruki Oshio).

ABSTRACT

The design of urban areas and building that utilizes the microclimatic effects of trees is a promising approach for reducing the severe heat stress caused by urban heat islands and global warming. Although trees can reduce heat stress through solar shading during the daytime, their influence on the air temperature under and around them during the nighttime, which is important for nighttime thermal comfort, has not yet been fully elucidated. In this study, we investigated the nocturnal cooling effect of trees in a physical urban space by the coupled numerical simulation of longwave radiative transfer and computational fluid dynamics. To represent the spatial structure of an actual urban space, airborne LiDAR-derived three-dimensional data of leaf area density distribution and building shape were used. The species-specific convective heat transfer coefficient was also considered. An analysis of the calculated sensible heat flux shows that both leaf area density and sky view factor are important factors in the production of cool air. According to the calculated distributions of air temperature and velocity, even under the condition of a certain degree of incident flow, the cooled air can flow down to the space under the crown, accumulate, and then diverge when the wind speed is sufficiently low in the crown owing to the crown drag. Buildings contribute to both the accumulation and dissipation of cool air. The findings of the present study suggest that cool spots can be produced during nighttime by trees planted near streets by devising a suitable arrangement and morphology of trees and buildings.

**1. Introduction**

The designing of urban areas and buildings utilizing the microclimatic effects of trees is a promising approach to alleviate the severe heat stress due to urban heat islands and global warming. In general, during summer daytime, trees can reduce human heat stress mainly by providing solar shading (Coutts et al., 2016; Kántor et al., 2018). Trees can block solar radiation without a significant increase in leaf surface temperature because of transpiration and the large convective heat transfer coefficient (CHTC) of leaves. The large value of the heat transfer coefficient of leaves is due to their size, which is relatively smaller than that of other urban objects. Many observational studies have investigated the outflow of cool air from an urban vegetated park to its vicinity at night (Spronken-Smith and Oke, 1998; Upmanis et al., 1998; Ca et al., 1998; Hamada and Ohta, 2010; Narita et al., 2011; Doick et al., 2014; Sugawara et al., 2016; Yan et al., 2018). However, the difference in the cooling effect between green morphologies, and that between trees and lawns, is not yet well understood.

During the night, if atmospheric radiative cooling is large, large convective heat transfer from the air to the leaf surface is expected owing to the large CHTC. Furthermore, an increase in leaf area index (LAI) results in a more negative sensible heat flux per unit ground area (heat flux from air to canopy) (Okada et al., 2016). From this point of view, trees have more cooling potential than ground cover plants. An increase in LAI reduces the radiative cooling of the leaves in the lower part of the crown, and these leaves block the gravitational flow of air cooled at the upper part of the crown. Therefore, the production of cool air by trees and its influence on the surrounding environment are considered complicated phenomena showing a close relationship among the convective heat exchange at the leaves, longwave radiative transfer, and airflow. These factors depend on the crown structure (density of leaves and its three-dimensional distribution) and the surrounding spatial geometry. The nocturnal cooling effect of trees cannot be fully elucidated based only on the observational studies because of the lack of sensitive and accurate measurement methods for such flow-radiation coupled phenomena in and around trees.

Numerical simulation is a feasible approach for investigating the flow of cool air produced by trees. Gross (1987) evaluated the influence of deforestation on the distribution of air temperature and velocity over a valley. Owing to the development of numerical simulation tools, recent studies have analyzed the variation in nocturnal air temperature and human heat stress indices according to the arrangement and morphology of buildings and vegetation (Ketterer and Matzarakis, 2014; Müller et al., 2014; Lee et al., 2016). However, studies on the flow of cool air produced by trees are lacking. Furthermore, previous relevant studies did not consider the structure of the crown. Moreover, there is lack of process knowledge on considering the effective arrangement and morphology of buildings and trees to utilize the nocturnal cooling effect of trees.

This study aimed to evaluate the potential of nocturnal cool air produced by trees under an ideal meteorological condition in a physical urban space. We performed a coupled simulation of longwave radiative transfer and computational fluid dynamics (CFD) to reveal the relationship between tree structure and sensible heat flux and elucidate the characteristics of cooled air flow produced by trees. The three-dimensional data of the distribution of leaf area density (LAD: total one-sided leaf area per unit volume) acquired through laser scanning (LiDAR: light detection and ranging) were used to represent the crown structure of urban trees in an actual scenario. Species-specific CHTCs were also considered. To study the effects of crown structure and CHTC in an actual urban environment, we investigated two popular urban tree species in Japan, namely, the Japanese zelkova (*Zelkova serrata*) and camphor laurel (*Cinnamomum camphora*).

## 2. Numerical simulation method

We developed a voxel-based coupled simulation program for modeling longwave radiative transfer and steady-state CFD. LAD distribution and building shape were represented by voxels (see Section 3.1 for data acquisition).

*2.1 Simulation procedure*

The leaf surface temperature, air temperature, and velocity were calculated using iterative methods. The simulation procedure consisted of two parts: radiative transfer simulation for calculating leaf surface temperature and steady-state non-isothermal CFD simulation for calculating air temperature and velocity. Both parts were interrelated: the calculated leaf surface temperature was used as input data for the CFD simulation, and the resulting air temperature and wind speed were used to calculate the new leaf surface temperature value in the radiative transfer simulation; and these processes were conducted iteratively. Figure 1 shows the flowchart of the numerical simulation process and the details are as follows:

(1) The view factor from the leaf surface to surrounding objects was calculated using the ray tracing method (see Section 2.3), and the initial velocity field was obtained using isothermal CFD simulation.

(2) The absorbed radiation was calculated using the view factor and initial surface temperature. Heat-balance equations of leaves (see Section 2.2) were solved to determine the leaf surface temperature by utilizing the initial wind speed, absorbed radiation, and initial air temperature. The calculations of absorbed radiation and leaf surface temperature were continued iteratively until a convergence of absorbed radiation was reached (i.e., the rate of change was less than 3% for all voxels).

(3) The air temperature and velocity were calculated using a non-isothermal CFD simulation (see Section 2.4) using the heat flux from leaves that was given based on the leaf surface temperature.

(4) Using the resulting air temperature and wind speed, iterative calculations of absorbed radiation and leaf surface temperature were conducted until a convergence of absorbed radiation was reached (similar to the above process (2)). Then, a non-isothermal CFD simulation was conducted using the resulting leaf surface temperature (similar to the above process (3)). These processes were continued iteratively until the convergence of air temperature was reached (the rate of change in air temperature was less than 1% for all voxels).

The surface temperatures of the buildings and ground were treated as boundary conditions and fixed at a prescribed value. For trees, only leaves were considered, but woody elements were not considered.

*2.2 Heat balance and absorbed radiation of leaves*

Transpiration and shortwave radiation can be neglected at night. The heat capacity of tree leaves is sufficiently small when discussing the microclimatic effects. The emissivity of a leaf surface is generally high (0.95–0.98) (Ribeiro da Luz and Crowley, 2007; López et al., 2012), and can thus be approximated to 1. Therefore, the heat balance of the leaves at night can be simplified as follows:

$$R_{abs} = H + \sigma T_l^4, \tag{1}$$

$$H = h_c(T_l - T_a), \tag{2}$$

where $R_{abs}$ is the absorbed radiation flux [W m$^{-2}$], $H$ is the sensible heat flux [W m$^{-2}$], $\sigma$ is the Stefan–Boltzmann constant (= 5.67×10$^{-8}$ W m$^{-2}$ K$^{-4}$), $T_l$ and $T_a$ are the temperatures of the leaf surface and air [K], respectively, and $h_c$ is the CHTC [W m$^{-2}$ K$^{-1}$]. Here, we considered both sides of a leaf; therefore, the total amount of absorbed radiation and sensible heat of a voxel are described as $2\rho_l V_{vox} R_{abs}$ and $2\rho_l V_{vox} H$, respectively, where $\rho_l$ is the LAD [m$^2$ m$^{-3}$] and $V_{vox}$ is the volume of the voxel [m$^3$]. The CHTC depends on the flow characteristics around the leaf surface and leaf size. We used the species-specific relationship between the CHTC and mean wind speed inside a voxel (see Section 3.1 for experimental conditions).

The absorbed radiation flux is denoted as

$$R_{abs} = F_s R_s + \sum_{i=1}^{N_g} F_{g,i} R_{g,i} + \sum_{i=1}^{N_b} F_{b,i} R_{b,i} + \sum_{i=1}^{N_l} F_{l,i} \sigma T_{l,i}^4, \tag{3}$$

where $F_s$, $F_{g,i}$, $F_{b,i}$, and $F_{l,i}$ are the view factors from the target leaf surface to the sky, the $i$th ground voxel, the $i$th building voxel, and leaf surface of the $i$th tree voxel, respectively [-], $R_s$ is the longwave radiation from the sky to the top of the urban canopy [W m$^{-2}$], $N_g$, $N_b$, and $N_l$ are the number of voxels of the ground, buildings, and trees, respectively, $R_{g,i}$ and $R_{b,i}$ are the radiant emittance of the $i$th ground and building voxels, respectively [W m$^{-2}$], and $T_{l,i}$ is the leaf surface temperature of the $i$th tree voxel [K].

*2.3 Calculation of view factor*

The view factor was calculated using the Monte Carlo ray tracing method. The ray emission position within a voxel was determined using random numbers. Then, the direction of the normal vector of a leaf surface was determined using random numbers; a probability distribution within a range of 0° to 90° was considered for the zenith angle (species-specific leaf inclination angle distribution), and the azimuth angle was assumed to follow a uniform distribution. The zenith angle ($\theta$) and azimuth angle ($\varphi$) within the spherical coordinate system (its zenith direction is the leaf normal direction) were determined using random numbers and a ray was emitted in this direction. The zenith angle was determined within a range of 0°–180° to represent the adaxial or abaxial side of the leaf. The weight of the ray ($w$) was initially set to $\cos\theta\sin\theta$, which highlights that the view factor from the differential area on the leaf ($dS_l$) to the differential area on the sphere in the direction ($\theta$, $\varphi$) is $\cos\theta\sin\theta d\theta d\varphi dS_l/\pi$. When a voxel with LAD > 0 was passed by the ray, the weight of the ray was changed to $Pw$ and a weight of $(1 - P)w$ was added to the voxel, where $P$ is the gap fraction of the voxel:

$$P = \exp(-G\rho_l d), \tag{4}$$

where $G$ is the mean projection area of a unit leaf area onto a plane perpendicular to the ray direction [-] and $d$ is the path length [m]. When the ray hits the ground, building, or boundary of the calculation domain (sky), the remaining weight is added to the object hit by the ray and the trace is terminated. A total of 2000 rays were emitted per voxel and the view factor was calculated as follows:

$$F_j = w_j/W, \tag{5}$$

where $F_j$ is the view factor from the leaf surface of the target voxel to object $j$ (sky, $i$th ground voxel, $i$th building voxel, or $i$th tree voxel) [-], $w_j$ is the weight of object $j$ [-], and $W$ is the total weight [-]. These calculations were conducted for all voxels with a LAD > 0. The view factor from a surface to the sky is generally defined between 0 and 1 (sky view factor: SVF); however, in our case, it ranged between 0 and 0.5 because both sides of each leaf were considered. Hereafter, the value of SVF is twice the calculated view factor from the surface to the sky so as to maintain consistency with other studies.

*2.4 Momentum transfer with tree models*

We conducted a steady-state CFD simulation to obtain the velocity and temperature fields. The mean flow was modeled using the Reynolds-averaged Navier-Stokes (RANS) equations with the standard k-epsilon turbulence model, and the buoyancy effect was considered in air density with the Boussinesq approximation. For steady flows in the atmospheric surface layer (i.e., neglecting the Coriolis force)—with these approximations—the basic transport equations of mass, mean momentum, temperature, and turbulent kinetic energy and its dissipation rate are given as

$$\frac{\partial \overline{u_i}}{\partial x_i} = 0, \tag{6}$$

$$\overline{u_j}\frac{\partial \overline{u_i}}{\partial x_j} = \frac{\partial}{\partial x_j}\left[(\nu + \nu_t)\left(\frac{\partial \overline{u_i}}{\partial x_j} + \frac{\partial \overline{u_j}}{\partial x_i}\right)\right] - \frac{2}{3}k\delta_{ij} - \frac{1}{\rho}\frac{\partial \overline{p}}{\partial x_i} + g_i\frac{\overline{T}-T_0}{T_0} + S_{u,i}, \tag{7}$$

$$\overline{u_j}\frac{\partial \overline{T}}{\partial x_j} = \frac{\partial}{\partial x_j}\left[\left(\frac{\nu}{\text{Pr}} + \frac{\nu_t}{\text{Pr}_t}\right)\frac{\partial \overline{T}}{\partial x_j}\right] + S_T, \tag{8}$$

$$\overline{u_j}\frac{\partial k}{\partial x_j} = \frac{\partial}{\partial x_j}\left[\left(\nu + \frac{\nu_t}{\sigma_k}\right)\frac{\partial k}{\partial x_j}\right] + P_k - \varepsilon + S_k, \tag{9}$$

$$\overline{u_j}\frac{\partial \varepsilon}{\partial x_j} = \frac{\partial}{\partial x_j}\left[\left(\nu + \frac{\nu_t}{\sigma_\varepsilon}\right)\frac{\partial \varepsilon}{\partial x_j}\right] + C_{\varepsilon 1}P_k\frac{\varepsilon}{k} - C_{\varepsilon 2}\frac{\varepsilon^2}{k} + S_\varepsilon, \tag{10}$$

where $\overline{u_i}$ is the mean velocity in $i$th direction [m s$^{-1}$], $\overline{p}$ is the mean hydrostatic pressure [Pa], $\overline{T}$ is the mean air temperature [K], $k$ is the turbulent kinetic energy (TKE) [m$^2$ s$^{-2}$], $\varepsilon$ is the dissipation rate of TKE [m$^2$ s$^{-3}$]. The turbulent viscosity of the standard k-epsilon model is $\nu_t = C_\mu k^2/\varepsilon$, and $P_k = \nu_t\left(\frac{\partial \overline{u_i}}{\partial x_j} + \frac{\partial \overline{u_j}}{\partial x_i}\right)\frac{\partial \overline{u_i}}{\partial x_j}$ denotes the TKE production rate. The last terms of the right-hand side of Eqs. (7)–(10) are the source or sink terms that express the effects of the tree (discussed later). $\delta$ is the Kronecker delta. The reference temperature $T_0$ was set to be equal to the inlet air temperature. Other variables and coefficients are constants in this study: the kinematic viscosity of air $\nu$ [m$^2$ s$^{-1}$], air density $\rho$ [kg m$^{-3}$], gravitational acceleration $g_i$ [m s$^{-2}$], Prandtl number (Pr = 0.9), turbulent Prandtl number (Pr$_t$ = 0.7), and turbulence model coefficients ($C_\mu$ = 0.09, $C_{\varepsilon 1}$ = 1.44, $C_{\varepsilon 2}$ = 1.92, $\sigma_k$ = 1, $\sigma_\varepsilon$ = 1.3). The transport equations (Eqs. (6)–(10)) were solved using the commercial finite-volume solver STREAM ver. 14 (Software Cradle Co., Ltd., Osaka, Japan), which is well validated with other CFD software (Architectural Institute of Japan 2016). The pressure-velocity coupling was implicitly solved using the SIMPLEC scheme (Ferziger and Peric, 2002). The convective terms were discretized using the second-order quadratic upwind (QUICK) scheme in a staggered grid system. The convergence criteria of the inner iterations (i.e., criteria for conservation of physical properties) were the scaled residual values of $10^{-6}$ for $\bar{u}_i$, $k$, and $\varepsilon$ and $10^{-4}$ for $\bar{T}$, and those for steady state were $10^{-4}$ for all variables. These values were within the range recommended by the Architectural Institute of Japan (AIJ) guidelines for urban flows (Tominaga et al., 2008; AIJ, 2016).

The drag effect of tree leaves was calculated from the LAD. The additional source or sink terms of the effect of tree leaves for the transfer equations are modeled as follows (Hiraoka 1989; Green 1992):

$$S_{u,i} = -C_d \rho_l \bar{u}_i U, \qquad (11)$$

$$S_k = C_d \rho_l U^3, \qquad (12)$$

$$S_\varepsilon = \frac{\varepsilon}{k} C_{\varepsilon 4} S_k, \qquad (13)$$

where $C_d$ is the drag coefficient (= 0.2), $U$ is the mean wind speed [m s$^{-1}$], and $C_{\varepsilon 4}$ is a model coefficient (= 1.8). Note that Katul et al. (2004) showed that the value of drag coefficient is between 0.1 and 0.3 for most vegetation (about 0.3 for crops and about 0.2 for trees). Equations (12) and (13) are simplified versions of Green's (1992) model (see details for Sanz (2003)). Although Eqs. (12) and (13) likely underestimate TKE, the mean velocity can be calculated accurately (Mochida et al., 2008). We only used TKE to calculate the mean velocity (adiabatic boundary condition was used for ground and building surfaces as described in Section 3.2, and CHTC of a leaf was estimated using mean wind speed); therefore, the underestimation of TKE hardly affected the results of the present study. The sensible heat flux from leaves (Eq. (2)) was added to the source term in Eq. (8); $S_T = 2\rho_l H/\rho c_p$, where $c_p$ is the specific heat capacity of air [J kg$^{-1}$ K$^{-1}$].

## 3. Data and processing

*3.1 Study sites and tree models*

We considered two study sites: (1) an experimental field where the species-specific CHTC was obtained and a simulation was conducted for isolated trees to facilitate the interpretation of the results for the main target site, and (2) a physical urban space (the main target site).

The relationship between the CHTC of leaves and the mean wind speed for *Z. serrata* and *C. camphora* was obtained in the experimental field with an open area of 8800 m$^2$ in the city of Miyoshi, Aichi Prefecture, Japan (35.1355° N, 137.1001° E). Although the relationship between the CHTC of an isolated plate (single leaf) and dimensionless numbers (Grashof number and Reynolds number) is available in the literature (Schuepp, 1993; Albrecht et al., 2020), we parameterized the CHTC experimentally to better represent the characteristics of our target species (leaf size and movement of a leaf with the flow). The CHTC was estimated using the method of simultaneous heat-balance equations by using the thermography of sunlit and shaded parts of a crown (similar to the methods of Horie et al., 2006). Measurements using thermography (Thermo GEAR G100; Nippon Avionics Co., Ltd., Yokohama, Japan) were conducted in the experimental field on September 12, 2012. The *Z. serrata* and *C. camphora*, with heights of 6.4 and 4.7 m, respectively (Fig. 2 left), were measured from four positions surrounding each tree between 0900 local standard time (LST = UTC + 9 h) to 1700 LST with an interval of 1 h. The wind speed was measured at a height of 5 m using a three-dimensional ultrasonic anemometer (Model 81000; R.M. Young Co., Michigan, USA) installed near the trees. The CHTC was obtained by solving simultaneous heat-balance equations for sunlit and shaded leaves. The details of the calculations are provided in Appendix A.

A simulation was conducted for isolated trees to compare the results with those for a physical urban space and investigate the influence of the difference in SVF according to the surrounding spatial geometry. We used terrestrial LiDAR-derived LAD data for the *Z. serrata* and *C. camphora* trees used to obtain the CHTC values (Oshio and Asawa, 2020) (Fig. 2, right). The voxel size was 0.3 m × 0.3 m × 0.3 m. The probability distribution of leaf inclination angle for each tree species was also derived from terrestrial LiDAR data and was used for the radiative transfer simulation. Oshio and Asawa (2020) have already provided details of the voxel data for these isolated trees. The main target site was Hisaya-Odori Street in the city of Nagoya, Aichi Prefecture, Japan (35.1709° N, 136.9085° E) (Fig. 3(a)). The width of this street is 110 m: a 10-m-wide sidewalk exists along the buildings on either side of the street, with a 10-m-wide roadway next to the sidewalk, and a 70-m-wide street park at the center of the street. The average building height is 40 m. This street is lined with numerous broad-leaved trees. The main species are *Z. serrata* and *C. camphora,* with heights of 10–14 m. We generated voxel data of 1 m × 1 m × 1 m (Fig. 3(b)(c)) using a voxel-based model of the LAD distribution and building shapes derived from airborne LiDAR data (Oshio et al., 2015; Oshio and Asawa, 2016). The airborne LiDAR data were acquired on September 6, 2010, during which time, there were leaf-on conditions in Japan. The probability distribution of the leaf inclination angle for each tree species, which was derived from the terrestrial LiDAR data over the site and used for calculating the LAD, was used for the radiative transfer simulation. The details on the voxel data have been provided by Oshio et al. (2015) and Oshio and Asawa (2016).

*3.2 Boundary conditions and simulation domain*

The meteorological conditions for the simulation were derived from observation data obtained from the Nagoya Local Meteorological Observatory, which is operated by the Japan Meteorological Agency, in the city of Nagoya, Aichi Prefecture, Japan (35.1682° N, 136.9649° E). To conduct a simulation for summer nighttime, the mean values of air temperature and wind speed were calculated using the data acquired at 2000 LST during August 2015. Only data acquired under calm (wind speed lower than 2 m s$^{-1}$) and clear-sky conditions were used to calculate the mean values because the cooling effect of vegetation is expected to occur under such conditions (Sugawara et al., 2016). Longwave radiation from the sky was not measured at the site; therefore, it was estimated using Brunt's formula with coefficients provided by Yamamoto (1950).

The computational domain size and boundary conditions of the CFD simulation were determined based on the AIJ guidelines (Tominaga et al., 2008; AIJ, 2016). The total domain size of the Hisaya-Odori Street simulation was 930 m × 630 m × 450 m in the streamwise, transverse, and vertical directions, respectively. It comprises an inner area of 180 m × 130 m × 60 m, where buildings and trees exist with a homogeneous grid spacing of 1 m. Outside the inner area, the grid was stretched toward the boundary of the simulation domain with a stretching ratio of 1.2 that was within the range of recommended value by the AIJ guidelines. The minimum mesh size (1 m) was also within the range of the recommended value (less than one-tenth of the characteristic length of the building). The inflow wind direction was northward along the Hisaya-Odori Street. Inlet vertical profiles for mean horizontal velocity $U(z)$, turbulence kinetic energy $k(z)$, and dissipation rate $\varepsilon(z)$, where $z$ is the vertical position above ground, were set as follows:

$$U(z) = U_r \left(\frac{z}{z_r}\right)^\alpha,$$

$$k(z) = [U(z)I(z)]^2 = \left[0.1 U(z) \left(\frac{z}{z_G}\right)^{-\alpha-0.05}\right]^2,$$

$$\varepsilon(z) = \sqrt{C_\mu} k(z) \frac{U_r}{z_r} \alpha \left(\frac{z}{z_r}\right)^{\alpha-1},$$

where $U_r$ is the wind speed at a reference height $z_r$ (here, 1.6 m s$^{-1}$ at a height of 17.8 m), $\alpha$ is a surface roughness parameter, and $z_G$ is the boundary layer height, both of which are determined based on the terrain category of the AIJ guidelines. In this study, these values were set according to category III (the area consists of low-rise buildings or sparse 4–9 story buildings: $z_G = 450$ m and $\alpha = 0.2$). $I(z)$ is the turbulent intensity and $C_\mu = 0.09$ is the model coefficient. The lateral and top boundaries of the domain were set to free-slip conditions, whereas the log-law wall function for smooth surfaces was applied to the ground and building surfaces. The adiabatic boundary condition was used. The temperature of the inflow was set as vertically uniform (= 27.9 °C, similar to the meteorological data). The initial leaf surface temperature was set to this temperature. We assumed that the surface temperature of the ground and buildings also equaled this temperature ($T_{a,met}$), and the radiant emittance was given by $\sigma T_{a,met}^4$ (constant value during the simulation). The settings of the CFD simulation for the isolated trees are described in Appendix B.

## 4. Results

*4.1 Convective heat transfer coefficient*

Figure 4 shows the relationship between CHTC and wind speed. We fitted a power function of the wind speed for each species and used it in the numerical simulation. The *Z. serrata* tree had a larger CHTC, probably because it has smaller leaves (i.e., smaller characteristic length). This is a natural result of heat-transfer theory. We compared the obtained CHTC with that derived using a semiempirical formula for an isolated ellipsoidal flat plate (Campbell and Norman, 1998). For the semiempirical formula, the average leaf width derived from the terrestrial LiDAR data was used as the characteristic leaf dimension (0.028 m and 0.034 m for *Z. serrata* and *C. camphora*, respectively). According to Fig. 4, the dependence of CHTC on wind speed was underestimated by our CHTC. This can be attributed to the shelter effect; that is, the wind speed over a leaf is decreased by the drag of leaves on the windward side; however, this does not occur for an isolated leaf. However, the difference in CHTC value between our result and the semiempirical formula was relatively small under low wind speed, which was assumed in the numerical simulation. The CHTC value for free convection was not obtained because most of the measurement data were obtained under conditions with a dominant forced convection. The CHTC value given by the power function was possibly smaller than the true value in the immediate vicinity of a wind speed of zero. However, this hardly affected the results of our numerical simulation because we assumed a condition with a certain degree of incident flow (1.6 m s$^{-1}$ above the tree canopy), and the wind speed for the inner part of the tree crown was expected as 0.2–0.3 m s$^{-1}$.

*4.2 Factors contributing to the negative sensible heat flux*

Figure 5 shows the variation in the sensible heat flux of a voxel (= $2\rho_l \Delta_{z\_vox} H$, where $\Delta_{z\_vox}$ is the height of a voxel) according to the LAD and SVF for the isolated trees. Hereafter, the term "sensible heat flux" means "sensible heat flux of a voxel" unless otherwise indicated. The negative sensible heat flux clearly increases with an increase in the LAD. Although a larger SVF offers a somewhat larger negative sensible heat flux, most of the variation in sensible heat flux depends on the LAD. To further interpret these results, the difference between the leaf surface temperature and air temperature (hereafter called $\Delta T \, (= T_l - T_a)$) was investigated. Figure 6 shows the variation in $\Delta T$ according to SVF and LAD. A larger SVF resulted in a smaller amount of absorbed radiation, yielding a lower leaf surface temperature (larger negative $\Delta T$). In comparison, $\Delta T$ changed toward zero with an increase in LAD. This is because a larger LAD results in a smaller SVF (the mutual radiation is dominant). The sensible heat flux is proportional to $\rho_l \times h_c \times \Delta T$. A larger negative sensible heat flux is caused by a larger LAD and larger negative $\Delta T$, but these conditions are incompatible. Figure 5 indicates that the LAD is a primary factor, meaning that the amount of cooling surface is important. This is in line with Okada et al. (2016), who experimentally showed that an increase in LAI yielded a smaller negative $\Delta T$ but larger negative sensible heat flux and lower air temperature.

Although the characteristics discussed in the previous paragraph are similar between the species, the negative sensible heat flux for *Z. serrata* was larger than that for *C. camphora* (Fig. 5). This is due to the difference in CHTC (*Z. serrata* has smaller leaves and larger CHTC). To confirm the influence of the difference in CHTC on the heat balance of a leaf, the variation in the sensible heat flux of leaf $H$ according to the absorbed radiation flux $R_{abs}$ and $\Delta T$ was investigated (Fig. 7). When observing each tree, the negative sensible heat flux of a leaf and negative $\Delta T$ decrease with an increase in absorbed radiation. In comparison, when comparing *Z. serrata* and *C. camphora* at a similar value of absorbed radiation, *Z. serrata* showed a smaller negative $\Delta T$ but larger negative sensible heat flux for a leaf than *C. camphora*. This means that owing to the larger CHTC, the leaves of *Z. serrata* absorb more heat (cool the air more) and maintain a higher surface temperature than those of *C. camphora*. These results highlight the importance of species-specific CHTC values when discussing the heat balance of leaves based on their surface temperature (e.g., estimating the sensible heat flux of leaves from thermal images).

The results obtained from the isolated trees revealed the basis of the negative sensible heat flux from a tree. Here, we discuss street trees in a physical urban space. Figure 8 shows the variation in the sensible heat flux according to the LAD and SVF for the trees in Hisaya-Odori Street. This tendency was significantly different from that observed for the isolated trees. The dependence of sensible heat flux on the LAD is unclear. When the data are divided into bins based on the SVF, the negative sensible heat flux increases clearly with increasing LAD, especially for bins with large SVF, indicating that both LAD and SVF are important factors. The difference between isolated trees and urban trees was in voxels with a small SVF ($< 0.3$). For such voxels, the increase in negative sensible heat flux with LAD was gentle and the correlation between sensible heat flux and LAD was low (Fig. 8). The small SVF was caused by the leaves in the upper part of the crown, surrounding trees, and surrounding buildings. Note that the voxel size was different between the isolated trees and urban trees. The influence of the difference in voxel size can be confirmed by focusing on the minimum sensible heat flux value for each LAD bin (e.g., size of 1 $m^2\ m^{-3}$). For a voxel size of 0.3 m (isolated trees), the minimum value decreases toward an LAD of 6 (Fig. 5). In contrast, for a voxel size of 1 m (Hisaya-Odori Street), the minimum value was almost stable for bins with LAD > 2 (Fig. 8). This can be explained by the fact that, when comparing voxels with different sizes but the same LAD, the larger voxels show a lower SVF because of mutual shading. Therefore, Fig. 5 does not indicate that the sensible heat flux generally continues to increase with increasing LAD. Overall, LAD is the primary factor for producing negative sensible heat flux for small, isolated trees with an entirely large SVF. Both LAD and SVF are important for large street trees in physical urban spaces with surrounding trees and buildings.

*4.3 Distribution of air temperature and velocity*

According to Section 4.2, the air is expected to be cooled significantly at the parts of the crown with large LAD and SVF values (e.g., the upper part of the crown). We confirmed that the cooled air can flow down to ground level under an ideal condition, where the incident flow is extremely weak, and there is no influence of surrounding trees and buildings (Appendix B). Here, the characteristics of cool air in actual conditions were investigated by analyzing the simulated distribution of air temperature and velocity for Hisaya-Odori Street. The horizontal sections of the distribution of the air temperature and velocity for different heights are shown in Fig. 9. For the windward side (southern part of the target domain), a region of air cooler than the ambient air can be seen in the tree crown (Fig. 9(a)). However, the cool air was advected horizontally and did not affect the space under the crown (Fig. 9(e)). Although the wind speed was decreased by the crown at a height of 12 m, the wind speed was high at a height of 1.5 m, where no tree crown existed, and the temperature was similar to that of the inflow air. Even at a height of 6 m, the wind speed did not decrease significantly. This is because lower part of the crown possessed more gaps (smaller number of voxels with leaves) (Fig. 3), especially for *C. camphora*.

At the leeward side, where the trees were planted more densely than those at the windward side, the wind speed decreased at a height of 6 m; this prevented the cool air from being dispersed. Then, the cool air accumulated in the space under the crown (Fig. 9(e)) and diverged (Fig. 9(f)). The difference in air temperature among spaces was large at a height of 12 m (i.e., a higher temperature was observed for the space between tree crowns). Further, the difference was small at heights of 6 and 1.5 m, and the air temperature was uniformly lower than that of inflow air by approximately 0.7 °C. The vertical section also clearly explains the aforementioned consideration (Fig. 10): the air temperature and wind speed were vertically uniform on the leeward side and a significant difference in air temperature was observed between the space under the crown on the windward side and that on the leeward side.

The influence of buildings on the airflow was observed as follows. At the windward side, side flows were observed along the buildings (northeastward flow from the sidewalk to the center of the street). The locally strong northeastward flow, which was seemingly caused by the complicated building geometry (the difference in height between buildings was large, as shown in Fig. 3), dispersed the cool air. For the leeward side comprising tall buildings with similar heights, a relatively strong northward flow was observed over the sidewalk. This flow appeared to be caused by the flow hitting the south-facing wall of the tall building after passing over the low-rise building on the windward side and may have disperse the cool air. In contrast, on the north side of the leeward side, the flow of cool air hit the building and split into northward and southward flows. This building seems to contribute to the accumulation of cool air.

**5. Discussion**

*5.1 Findings of the present study and their relation to previous studies on nocturnal cooling effect of urban trees*

The park breeze, which is the divergent flow of cool air due to the pressure gradient between the colder park and the warmer vicinity, has attracted attention as the cooling effect of vegetation during nighttime (Oke, 1989; Eliasson and Upmanis, 2000; Narita et al., 2011; Sugawara et al., 2016). This park breeze occurs under calm conditions (weak incident flow above the canopy: wind speed lower than 1–1.5 m s$^{-1}$) (Narita et al., 2011; Sugawara et al., 2016). The park breeze has been observed not only for parks mainly consisting of lawn fields but also for those covered by trees, although it is generally considered that an open lawn field with a large SVF plays an important role in the park breeze. The influence of trees on park breezes has not been fully elucidated. The present study shows that cool air accumulation and divergent flow could occur in an urban street park consisting of trees. Even when there is a certain degree of inflow wind at the canopy height, the air cooled at the upper part of the tree crown can flow down to the space under the crown (human height) when the wind speed is sufficiently low in the crown owing to the crown drag. It is important that trees provide solar shading during the daytime and can decrease the air temperature of the space under the crown during nighttime. For urban parks, the distance between the front of cool airflow and the park is important for evaluating the microclimatic effects of urban parks during nighttime. In our case, the cool airflow is not expected to reach far from the trees; however, cool spaces under or near the street trees are usable at night.

The park breeze has been observed for urban parks with a certain degree of vegetated area ($\geq$ ~0.1 km$^2$). With respect to the influence of topography, Gross (1987) showed that a valley covered by trees can produce a larger amount of cool air than a valley without trees (after deforestation) by numerical simulation. Even for a smaller spatial scale, a cool downslope flow with low turbulence intensity was observed for a tree-planted garden on the stepped facade of a 15-story building (Hagishima et al., 2004; Hagishima, 2018). According to the results of the present study, the accumulation and divergence of cool air can occur even in a flat urban street park with a width of 70–80 m (the length was 180 m for our simulation). We suggest that cool spots can be produced in urban streets during the nighttime using trees by devising the arrangement of trees and buildings, building shape, and tree morphology (LAD distribution).

Both the LAD and SVF are important factors for producing a negative sensible heat flux. A large LAD is also important for decreasing the wind speed in the crown. Building utilization seems to be effective in blocking the incident flow and trapping cool air. However, a large LAD is incompatible with a large SVF, and buildings near trees decrease the SVF of the trees. Numerical simulation is a feasible way to explore urban design while considering the balance between these factors. The size, shape, and species of trees should be considered based not only on the thermal environment but also on other factors such as the visual aspect and brightness of the space. Our method, which is based on the three-dimensional distribution of LAD, can be used to evaluate the thermal environment by considering the differences between trees. Figures 5 and 8 indicate that the vertically integrated leaf area (LAI) seems to be important for explaining the negative sensible heat flux. However, according to Figs. 10 and 11, wind attenuation is required to bring cool air down, indicating that the integrated leaf area along the flow is also important. These results highlight the effectiveness of the three-dimensional LAD distribution for understanding the nocturnal cooling effect of trees. Our approach can also be used to reveal how trees contribute to park breezes.

*5.2 Validity and limitation of the present study*

While LAD distribution and longwave radiative transfer within tree crowns were considered in the simulation, some simplifications were made, such as setting the surface temperature to the equivalent of air temperature and adiabatic boundary condition for ground and building surfaces. Such simplification has been used to evaluate the potential of urban trees in improving the thermal environment (e.g., Manickathan et al., 2018). Given the nature of the present study, which is a numerical experiment to evaluate the nocturnal cooling potential of trees and to investigate factors related to tree structure and surrounding spatial structure that affect the cooling potential (not to reproduce results of specific field measurements), such a simplification does not deviate from the general process in this research field.

Furthermore, for the area where the cool air accumulated, setting the surface temperature to the same value as the air temperature did not significantly deviate from the actual condition because the trees intercepted incident solar radiation in the daytime (the area was covered by trees as shown in Fig. 3(a)), and the buildings blocked the westering sun. Previous field measurements reported that the air temperature inside the leafy forest (within and under the crown) was lower than that outside (above) the forest in the nighttime, and the vertical distribution of air temperature inside the forest was small (Finco et al., 2018; Kondo et al., 2016; Nölscher et al., 2016; Schilperoort et al., 2020). The difference in air temperature between the inside and outside of the forest increased with decreasing wind speed (Kondo et al., 2016; Schilperoort et al., 2020). The results of our numerical simulation are consistent with the general characteristics of the air temperature in tree-covered areas. The present study seeks to elucidate physical phenomena (behavior of cool air). Along with improving the method described below, we will apply our method to the evaluation of thermal comfort in future studies.

If the temperature of the incident air flow into the tree canopy increases after setting the ground surface temperature higher, the air needs to pass through the canopy for a longer distance to be cooled, similar to the results of the present study. In this case, the position where the cool air starts to flow down shifts to the leeward side. Meanwhile, the divergent flow of cool air gets heated when passing the ground surface, which is at a high temperature, and the temperature of the flow gradually approaches the ambient urban temperature (Narita et al., 2011). Therefore, an improvement in the treatment of ground surface temperature is required to investigate the effect of cool air on the thermal environment of surrounding areas and the difference in cooling effect among various arrangements and morphology of trees. In the case of areas with sparsely planted trees, the ground surface with a high temperature is also expected to affect the balance of longwave radiation of trees.

In addition, woody elements were not considered in the simulation. Although the influence of woody elements on the heat balance of trees is not negligible for trees with a small LAI, such as savanna trees (Kobayashi et al., 2012), the influence is expected to be small in our case because of the large LAI. The woody element treatment will be improved to apply our method to various tree species and to investigate the impact of tree maintenance, such as pruning, on the cooling effect. We also note that steady RANS models have difficulties in predicting the flows in the wake regions behind buildings and that the turbulence dissipation term $S_\varepsilon$ includes large uncertainty. Therefore, although our method is sufficient to capture the fundamentals of flow-radiation coupling phenomena occurring at night, the detailed values of dimensions, such as the extent of cooled air, which is important to landscape and urban design, may include non-negligible errors. Validation works based on observations in real urban areas and/or further simulations with highly accurate turbulence models should be conducted. Recently, large-eddy simulations considering individual trees have been performed (Matsuda et al., 2018; Grylls and Reeuwijk, 2021). Coupling the approach based on the LAD distribution with large-eddy simulations is an effective measure for validation.

## 6. Conclusions

The nocturnal cooling effect of urban street trees was investigated using a coupled simulation of longwave radiative transfer and CFD. We used voxel-based three-dimensional tree models considering the distribution of the LAD and species-specific CHTC. The numerical simulation was conducted under the conditions of a calm summer night.

Analysis of the sensible heat flux showed that LAD is a primary factor for producing negative sensible heat flux for small isolated trees with a large SVF. Both LAD and SVF are important factors for large street trees in physical urban spaces where the SVF becomes small because of the upper leaves and surrounding trees and buildings. A large LAD is incompatible with a large SVF: the increase in negative sensible heat flux with increasing LAD stopped at around LAD of 2 m$^{-2}$ m$^{-3}$ with a voxel size of 1 m × 1 m × 1 m.

According to the simulated distributions of air temperature and velocity, the air was cooled at the upper part of the crown and the wind speed decreased in the inner part of the crown. However, for the trees lined up in a single row, the wind speed was still high for the lower part of the crown and the space under the crown, resulting in the dissipation of cool air along the prevailing flow. In contrast, for the area covered by trees, the wind speed decreased within the entire crown and the cool air flowed down to ground level. We showed that the accumulation and divergent flow of cool air can occur even in a flat urban street park consisting of trees (without lawn fields) with a width of several tens of meters.

The primary factors governing the nocturnal accumulation and divergence of cool air are (1) the amount of leaves (LAD) to produce large negative sensible heat flux, (2) wind attenuation to bring cool air down to ground (human height) level, and (3) maintaining sufficient distance from surrounding buildings and trees to obtain a large enough SVF. Factors (2) and (3) are contradictory, so we have to consider the balance between them when designing greenery spaces in an actual urban environment. For this purpose, detailed three-dimensional simulations, such as ours, are a promising approach. Our methods and results will contribute to future landscape designs, providing a way to determine how dense trees should be planted (in other words, how large a park should be) and what kind of plant species should be chosen as urban greenery.

**CRediT authorship contribution statement**

**Haruki Oshio:** Conceptualization, Methodology, Software, Validation, Formal analysis, Investigation, Writing - original draft, Writing - review & editing, Visualization. **Tomoki Kiyono:** Conceptualization, Methodology, Software, Validation, Formal analysis, Investigation, Writing - review & editing. **Takashi Asawa:** Conceptualization, Resources, Data curation, Writing - review & editing, Supervision, Project administration.

**Declaration of competing interest**

The authors declare no competing interests.

**Acknowledgement**

The authors thank Mr. Tomoki Ishimaru for his assistance.

**Appendix A. Calculation of convective heat transfer coefficient**

The heat-balance equation of a leaf during daytime is written as

$$\alpha(G_s S_{dir} + S_{dif} + S_{ref}) + \varepsilon_l(L_d + L_u) = lE + 2h_c(T_l - T_a) + 2\varepsilon_l \sigma T_l^4, \tag{A1}$$

where $\alpha$ is the leaf absorptance for shortwave radiation [-], $G_s$ is the projection area of the unit leaf area on a plane perpendicular to the direct solar beam [-], $S_{dir}$ is the direct solar radiation [W m$^{-2}$], $S_{dif}$ is the sky-diffuse solar radiation [W m$^{-2}$], $S_{ref}$ is the reflected solar radiation [W m$^{-2}$], $L_d$ and $L_u$ are the downward and upward longwave radiation [W m$^{-2}$], respectively, $l$ is the latent heat of vaporization [J mol$^{-1}$], $E$ is the transpiration rate [mol m$^{-2}$ s$^{-1}$], and $\varepsilon_l$ is the emissivity of the leaf surface. When a sunlit leaf and a shaded leaf with the highest ($T_{l\_max}$) and the lowest ($T_{l\_min}$) surface temperature within the crown, respectively, are selected, the CHTC can be determined from heat-balance equations for the sunlit and shaded leaves as Eq. (A2) by assuming the following: (1) $S_{dir}$ equals zero for the shaded leaf; (2) $S_{dif}$, $S_{ref}$, $L_d$, $L_u$, $lE$, $h_c$, and $T_a$ are similar between the sunlit and shaded leaves; (3) $G_s$ equals one.

$$h_c = \{0.5\alpha S_{dir} - \varepsilon_l \sigma (T_{l\_max}^4 - T_{l\_min}^4)\}/(T_{l\_max} - T_{l\_min}). \tag{A2}$$

**Appendix B. Computational fluid dynamics simulation for isolated trees**

We conducted numerical simulations for the isolated trees assuming extremely calm conditions to investigate whether a downflow of cool air occurs under ideal conditions (weak incident flow and no influence of surrounding trees and buildings). The temperature and wind speed of inflow were 29.4 °C and 0.1 m s$^{-1}$, respectively (vertically uniform). The air temperature and wind speed were different from those of Hisaya-Odori Street because meteorological data for a day with extremely calm conditions (not the average for a specific period) were used. Other settings were basically the same as those for Hisaya-Odori Street, except that the total domain size was 120 m × 60 m × 30 m and the grid size of the inner area was 0.3 m. Figure A1 shows a vertical section of the calculated air temperature and velocity. The downflow of the cooled air is clearly observed. The difference in temperature between the inflow air and the downflow reached 0.6–0.8 °C and the wind speed of the downflow was 0.2–0.3 m s$^{-1}$. The difference in air temperature and velocity between *Z. serrata* and *C. camphora* can be attributed to the differences in CHTC and crown structure.

**References**

Albrecht, H., Fiorani, F., Pieruschka, R., Müller-Linow, M., Jedmowski, C., Schreiber, L., Schurr, U., Rascher, U., 2020. Quantitative estimation of leaf heat transfer coefficients by active thermography at varying boundary layer conditions. Frontiers in Plant Science 10, 1684.

Architectural Institute of Japan, 2016. AIJ benchmarks for validation of CFD simulations applied to pedestrian wind environment around buildings, Architectural Institute of Japan, Tokyo.

Ca, V.T., Asaeda, T., Abu, E.M., 1998. Reductions in air conditioning energy caused by a nearby park. Energy and Buildings 29, 83–92.

Campbell, G.S., Norman, J.M., 1998. An introduction to environmental biophysics, 2nd edition, Springer-Verlag, New York.

Coutts, A.M., White, E.C., Tapper, N.J., Beringer, J., Livesley, S.J., 2016. Temperature and human thermal comfort effects of street trees across three contrasting street canyon environments. Theoretical and Applied Climatology 124, 55–68.

Doick, K.J., Peace, A., Hutchings, T.R., 2014. The role of one large greenspace in mitigating London's nocturnal urban heat island. Science of the Total Environment 493, 662–671.

Eliasson, I., Upmanis, H., 2000. Nocturnal airflow from urban parks-implications for city ventilation. Theoretical and Applied Climatology 66, 95–107.

Ferziger, J.H., Peric, M., 2002. Computational Methods for Fluid Dynamics, third ed., Springer, Berlin.

Finco, A., Coyle, M., Nemitz, E., Marzuoli, R., Chiesa, M., Loubet, B., Fares, S., Diaz-Pines, E., Gasche, R., Gerosa, G., 2018. Characterization of ozone deposition to a mixed oak–hornbeam forest – flux measurements at five levels above and inside the canopy and their interactions with nitric oxide. Atmospheric Chemistry and Physics 18, 17945–17961.

Green, S.R., 1992. Modelling turbulent air flow in a stand of widely-spaced trees. PHOENICS Journal Computational Fluid Dynamics and its Applications 5, 294–312.

Gross, G., 1987. Some effects of deforestation on nocturnal drainage flow and local climate—A numerical study. Boundary-Layer Meteorology 38, 315–337.

Grylls, T., Reeuwijk, M., 2021. Tree model with drag, transpiration, shading and deposition: Identification of cooling regimes and large-eddy simulation. Agricultural and Forest Meteorology 298–299, 108288.

Kántor, N., Chen, L., Gál, C.V., 2018. Human-biometeorological significance of shading in urban public spaces - Summertime measurements in Pécs, Hungary. Landscape and Urban Planning 170, 241–255.

Katul, G.G., Mahrt, L., Poggi, D., Sanz, C., 2004. One- and two-equation models for canopy turbulence. Boundary-Layer Meteorology 113, 81–109.

Ketterer, C., Matzarakis, A., 2014. Human-biometeorological assessment of heat stress reduction by replanning measures in Stuttgart, Germany. Landscape and Urban Planning 122, 78–88.

Kobayashi, H., Baldocchi, D.D., Ryu, Y., Chen, Q., Ma, S., Osuna, J.L., Ustin, S.L., 2012. Modeling energy and carbon fluxes in a heterogeneous oak woodland: A three-dimensional approach. Agricultural and Forest Meteorology 152, 83–100.

Kondo, J., Sugawara, H., Naito, G., Hagiwara, S., 2016. Air temperature in National Park for Nature Study (NO 2). Miscellaneous Reports of the Institute for Nature Study 47, 1–22.

Hagishima, A., 2018. Green infrastructure and urban sustainability. AIP Conference Proceedings 1931, 020002.

Hagishima, A., Narita, K., Tanimoto, J., Misaka, I., Matsushima, A., Onoue, M., 2014. Field measurement on the micro climate around the building with the large stepped roof garden. Journal of Environmental Engineering (Transactions of AIJ) 69, 47–54.

Hamada, S., Ohta, T., 2010. Seasonal variations in the cooling effect of urban green areas on surrounding urban areas. Urban Forestry and Urban Greening 9, 15–24.

Hiraoka, H., Maruyama, T., Nakamura, Y., Katsura, J., 1989. A study on modelling of turbulent flows within plant and urban canopies: Formalization of turbulence model (Part 1). Journal of Architecture, Planning and Environmental Engineering (Transactions of AIJ) 406, 1–9.

Horie, T., Matsuura, S., Takai, T., Kuwasaki, K., Ohsumi, A., Shiraiwa, T., 2006. Genotypic difference in canopy diffusive conductance measured by a new remote-sensing method and its association with the difference in rice yield potential. Plant, Cell and Environment 29, 653–660.

Lee, H., Mayer, H., Chen, L., 2016. Contribution of trees and grasslands to the mitigation of human heat stress in a residential district of Freiburg, Southwest Germany. Landscape and Urban Planning 148, 37–50.

López, A., Molina-Aiz, F.D., Valera, D.L., Peña, A., 2012. Determining the emissivity of the leaves of nine horticultural crops by means of infrared thermography. Scientia Horticulturae 137, 49–58.

Manickathan, L., Defraeye, T., Allegrini, J., Derome, D., Carmeliet, J., 2018. Parametric study of the influence of environmental factors and tree properties on the transpirative cooling effect of trees. Agricultural and Forest Meteorology 248, 259–274.

Matsuda, K., Onishi, R., Takahashi, K., 2018. Tree-crown-resolving large-eddy simulation coupled with three-dimensional radiative transfer model. Journal of Wind Engineering and Industrial Aerodynamics 173, 53–66.

Mochida, A., Tabata, Y., Iwata, T., Yoshino, H., 2008. Examining tree canopy models for CFD prediction of wind environment at pedestrian level. Journal of Wind Engineering and Industrial Aerodynamics 96, 1667–1677.

Müller, N., Kuttler, W., Barlag, A.-B., 2014. Counteracting urban climate change: Adaptation measures and their effect on thermal comfort. Theoretical and Applied Climatology 115, 243–257.

Narita, K., Sugawara, H., Yokoyama, H., Misaka, I., Matsushima, D., 2011. Field measurements on the cooling effect of the imperial palace and its thermal influence on the surrounding built up area. Journal of Environmental Engineering (Transactions of AIJ) 76, 705–713.

Nölscher, A.C., Yañez-Serrano, A.M., Wolff, S., Carioca de Araujo, A., Lavrič, J.V., Kesselmeier, J., Williams, J., 2016. Unexpected seasonality in quantity and composition of Amazon rainforest air reactivity. Nature Communications 7, 10383.

Okada, M., Okada, M., Kusaka, H., 2016. Dependence of atmospheric cooling by vegetation on canopy surface area during radiative cooling at night: Physical model evaluation using a polyethylene chamber. Journal of Agricultural Meteorology 72, 20–28.

Oke, T.R., 1989. The micrometeorology of the urban forest. Philosophical Transactions of the Royal Society of London Series B 324, 335–349.

Oshio, H., Asawa, T., 2016. Estimating the solar transmittance of urban trees using airborne LiDAR and radiative transfer simulation. IEEE Transactions on Geoscience and Remote Sensing 54, 5483–5492.

Oshio, H., Asawa, T., 2020. Verifying the accuracy of the leaf area density distribution of an individual tree derived from terrestrial laser scanning while considering the penetration of beams into the crown and the influence of wind. Journal of the Remote Sensing Society of Japan 40, S34–S43.

Oshio, H., Asawa, T., Hoyano, A., Miyasaka, S., 2015. Estimation of the leaf area density distribution of individual trees using high-resolution and multi-return airborne LiDAR data. Remote Sensing of Environment 166, 116–125.

Ribeiro da Luz, B., Crowley, J.K., 2007. Spectral reflectance and emissivity features of broad leaf plants: Prospects for remote sensing in the thermal infrared (8.0–14.0 μm). Remote Sensing of Environment 109, 393–405.

Sanz, C., 2003. A note on k - ε modelling of vegetation canopy air-flows. Boundary-Layer Meteorology 108, 191–197.

Schilperoort, B., Coenders-Gerrits, M., Rodríguez, C.J., van der Tol, C., van de Wiel, B., Savenije, H., 2020. Decoupling of a Douglas fir canopy: a look into the subcanopy with continuous vertical temperature profiles. Biogeosciences 17, 6423–6439.

Schuepp, P.H., 1993. Tansley Review No. 59 Leaf boundary layers. New Phytologist 125, 477–507.

Spronken-Smith, R.A., Oke, T.R., 1998. The thermal regime of urban parks in two cities with different summer climates. International Journal of Remote Sensing 19, 2085–2104.

Sugawara, H., Shimizu, S., Takahashi, H., Hagiwara, S., Narita, K., Mikami, T., Hirano, T., 2016. Thermal influence of a large green space on a hot urban environment. Journal of Environmental Quality 45, 125–133.

Tominaga, Y., Mochida, A., Yoshie, R., Kataoka, H., Nozu, T., Yoshikawa, M., Shirasawa, T., 2008. AIJ guidelines for practical applications of CFD to pedestrian wind environment around buildings. Journal of Wind Engineering and Industrial Aerodynamics 96, 1749–1761.

Upmanis, H., Eliasson, I., Lindqvist, S., 1998. The influence of green areas on nocturnal temperatures in a high latitude city (Göteborg, Sweden). International Journal of Climatology 18, 681–700.

Yamamoto, G., 1950. On nocturnal radiation Part II. Numerical calculation. Journal of the Meteorological Society of Japan Ser. II 28, 11–20.

Yan, H., Wu, F., Dong, L., 2018. Influence of a large urban park on the local urban thermal environment. Science of the Total Environment 622–623, 882–891.

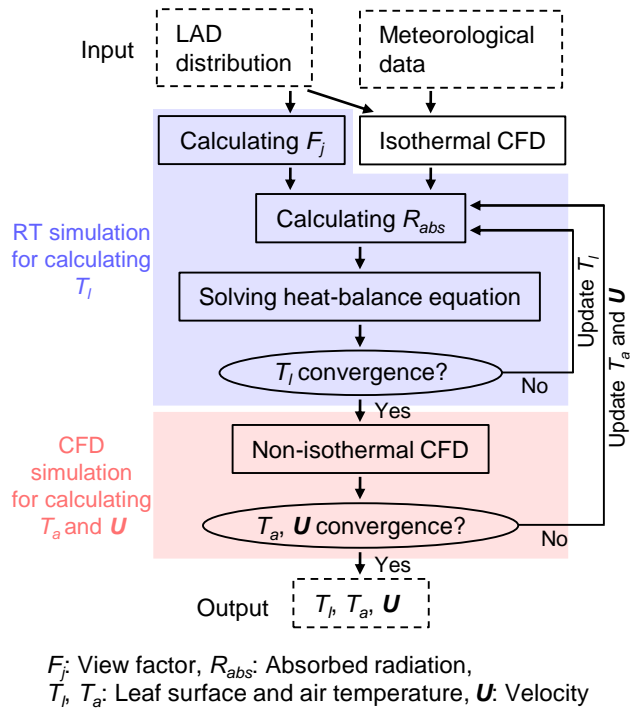

Fig. 1. Flowchart of the coupled simulation of longwave radiative transfer (RT) and CFD.

(a) *Z. serrata*

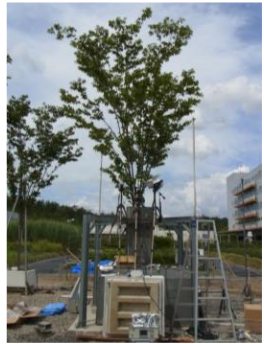 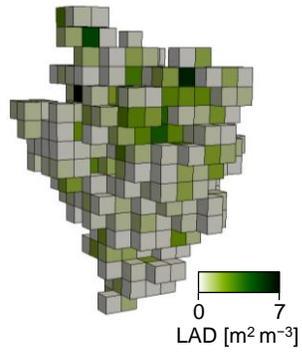

(b) *C. camphora*

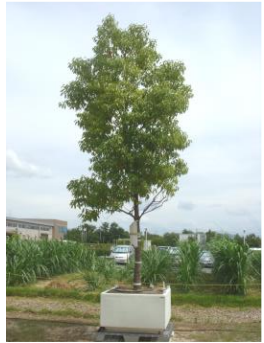 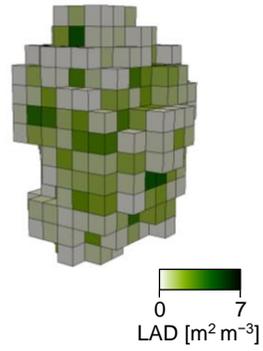

Fig. 2. Photograph (left) and voxel model (right) of isolated trees: (a) *Z. serrata*; (b) *C. camphora*.

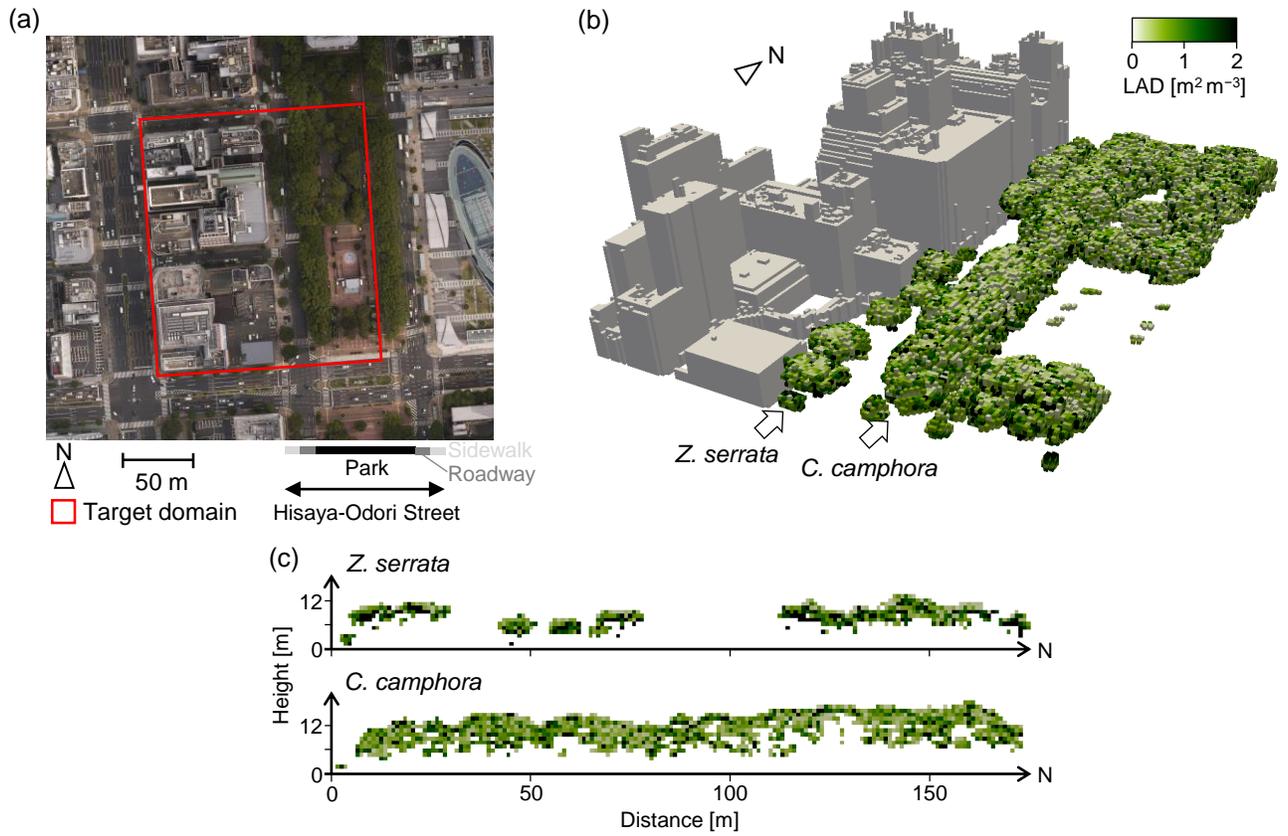

Fig. 3. Target site and three-dimensional spatial data: (a) aerial photograph of the target site; (b) airborne LiDAR-derived voxel model of the trees and buildings; (c) vertical section of the tree model.

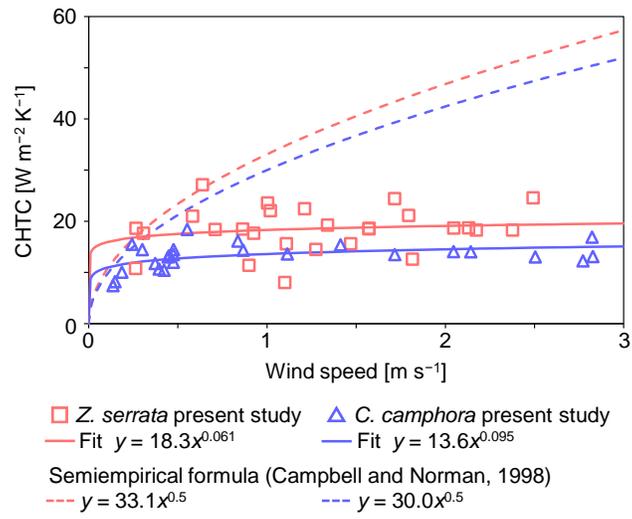

Fig. 4. Relationship between CHTC of a leaf and wind speed.

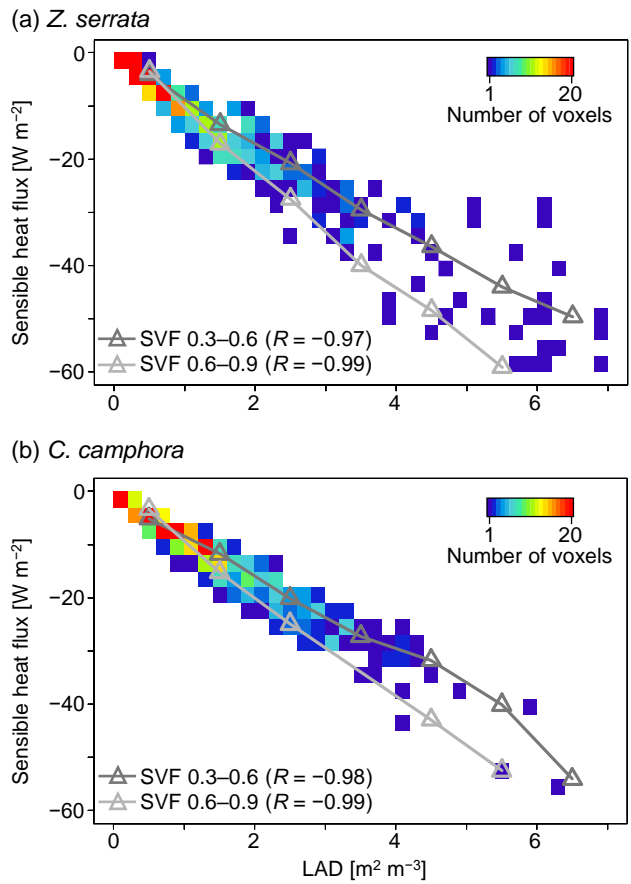

Fig. 5. Relationship between the sensible heat flux of a voxel and LAD for the isolated trees: (a) *Z. serrata*; (b) *C. camphora*. The color indicates the data count (number of voxels) within a bin of sensible heat flux and LAD. The triangle symbol shows the mean sensible heat flux value within a bin of LAD (1 m$^2$ m$^{-3}$ in size) for the data binned according to the sky view factor (SVF). The correlation coefficient ($R$) between sensible heat flux and LAD for each SVF bin is also shown.

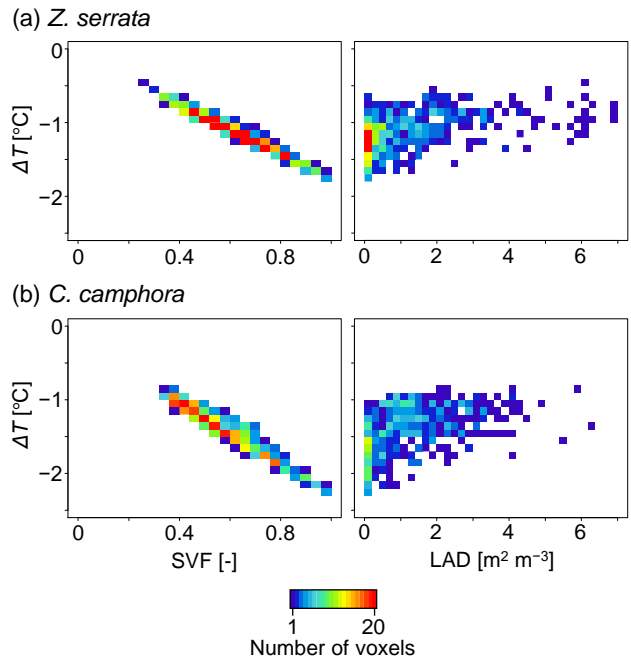

Fig. 6. Relationship between the difference between leaf surface temperature ($T_l$) and air temperature ($T_a$) ($\Delta T = T_l - T_a$) and the variables related to tree structure (sky view factor (SVF) and LAD) for the isolated trees: (a) *Z. serrata*; (b) *C. camphora*. The color indicates the data count (number of voxels) within a bin of $\Delta T$ and variables.

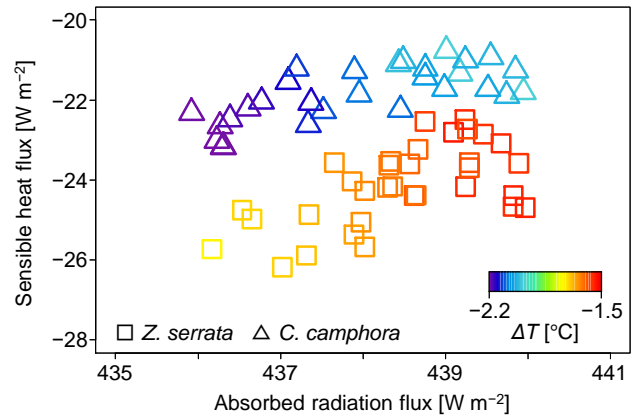

Fig. 7. Relationship between the sensible heat flux of a leaf surface, absorbed radiation flux of a leaf surface, and difference between leaf surface temperature ($T_l$) and air temperature ($T_a$) ($\Delta T = T_l - T_a$) for the isolated trees. Only data with a small value of absorbed radiation flux (< 440 W m$^{-2}$) are shown.

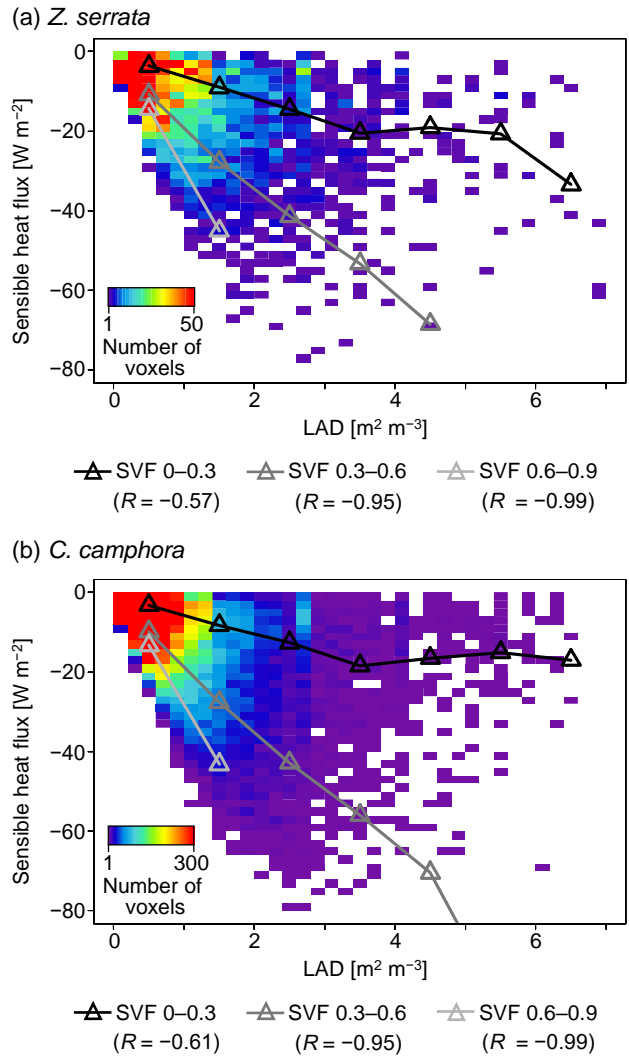

Fig. 8. Similar to Fig. 5 but for the trees in Hisaya-Odori Street.

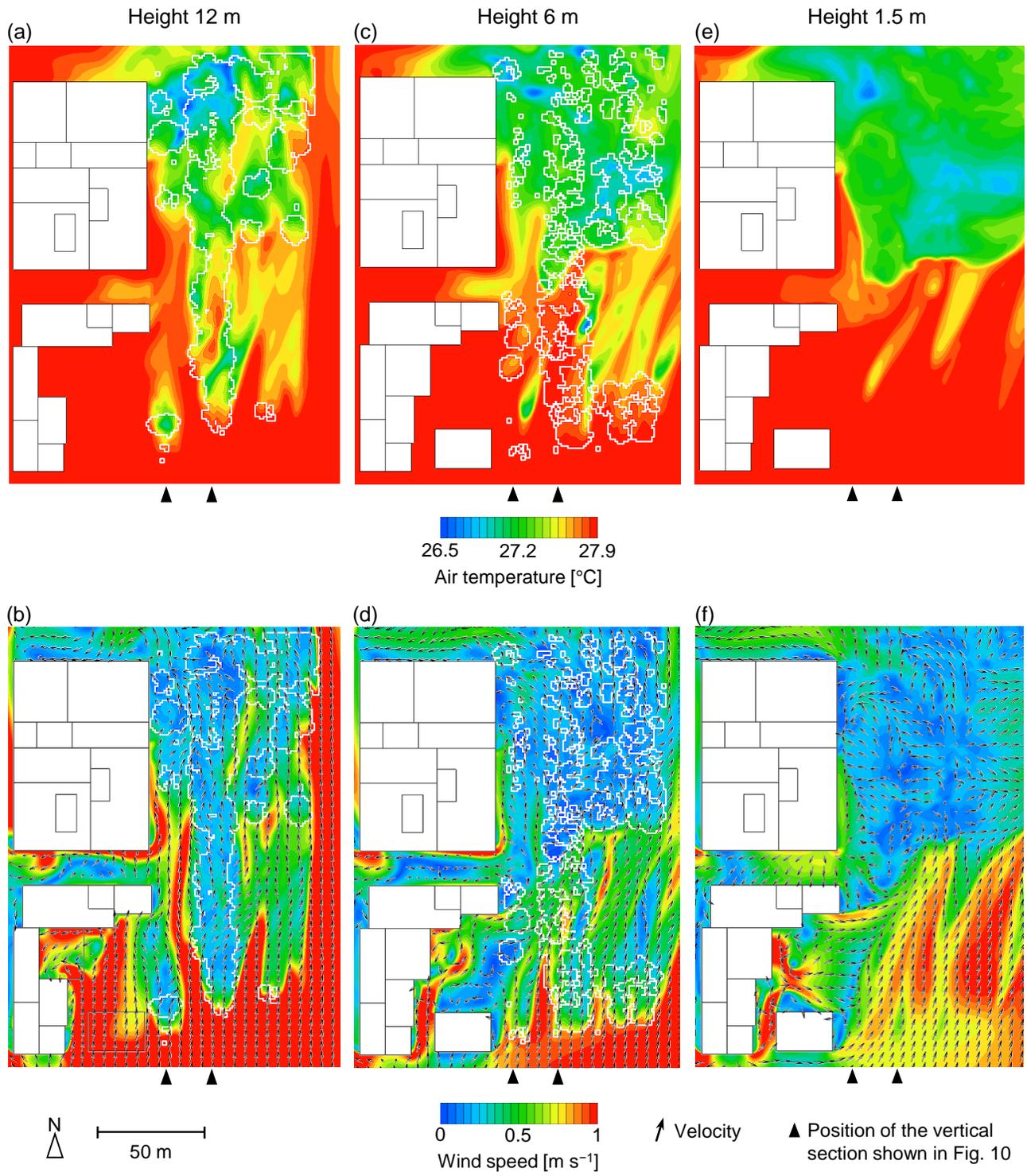

Fig. 9. Horizontal section of the simulated (a, c, e) air temperature and (b, d, f) velocity. Results for the different levels of height from the ground are shown: (a, b) 12 m; (c, d) 6 m; (e, f) 1.5 m.

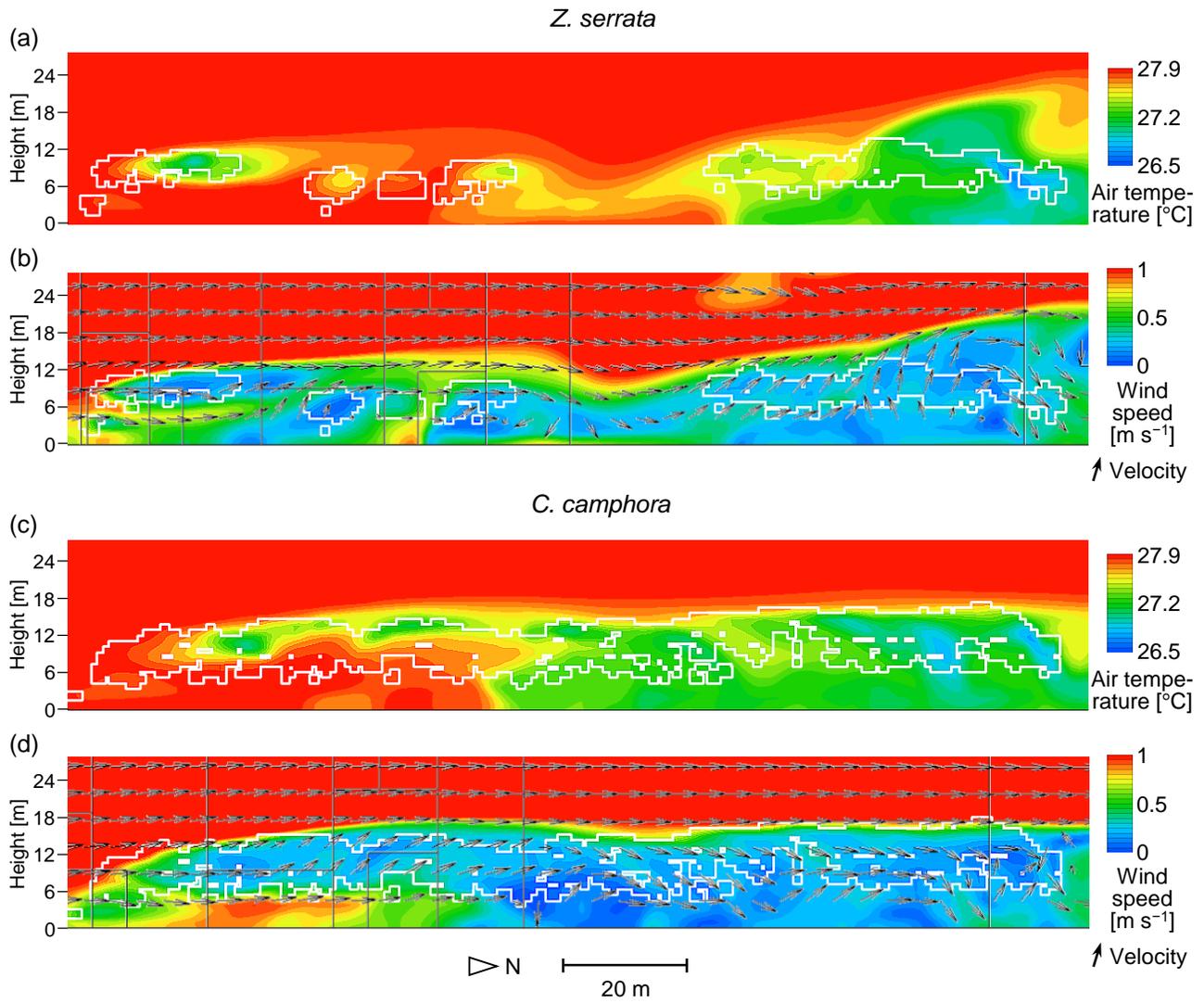

Fig. 10. Vertical section of the simulated (a, c) air temperature and (b, d) velocity: (a, b) *Z. serrata*; (c, d) *C. camphora*.

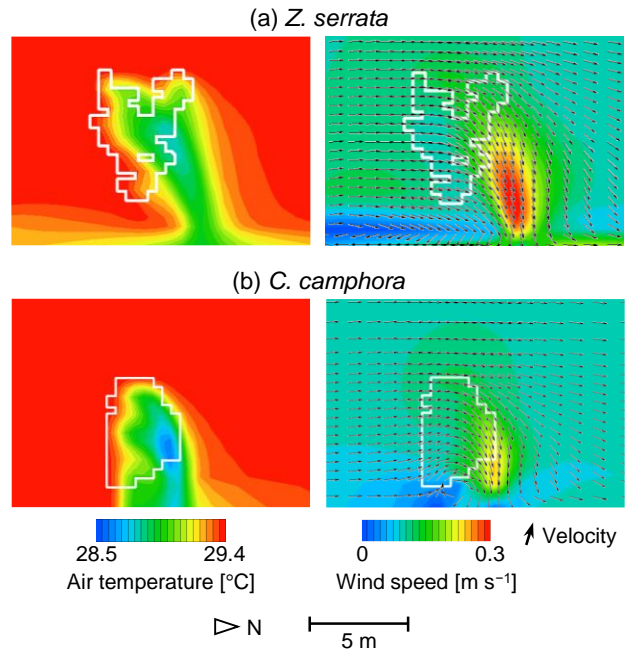

Fig. A1. Vertical section of the simulated air temperature (left) and velocity (right) for the isolated trees: (a) *Z. serrata*; (b) *C. camphora*.